\documentclass[particles,article,accept,pdftex,moreauthors]{Definitions/mdpi}
\usepackage{xcolor}
\firstpage{1}
\pubvolume{1}
\issuenum{1}
\articlenumber{0}
\pubyear{2023}
\copyrightyear{2023}
\datereceived{ }
\daterevised{ } 
\dateaccepted{ }
\datepublished{ }
\hreflink{https://doi.org/} 

\Title{Hyperon Production in Bi+Bi Collisions at NICA and Angular Dependence of Hyperon Spin Polarization}

\TitleCitation{Hyperon production in Bi+Bi collisions at NICA and angular dependence of hyperon spin polarization}

\Author{Nikita S. Tsegelnik$^{1}$\orcidN{}, Vadym Voronyuk$^{2}$\orcidV{} and Evgeni E. Kolomeitsev$^{3}$\orcidA{}*}

\AuthorNames{Nikita S. Tsegelnik , Vadym Voronyuk, and Evgeni E. Kolomeitsev}

\AuthorCitation{ Tsegelnik, N.S.; Voronyuk, V.; Kolomeitsev, E.E.}

\address{%
$^{1}$ \quad Laboratory of Theoretical Physics, Joint Institute for Nuclear Research, RU-141980 Dubna, Russia; tsegelnik@jinr.ru\\
$^{2}$ \quad Laboratory of High Energy Physics, Joint Institute for Nuclear Research, RU-141980 Dubna, Russia; Vadim.Voronyuk@jinr.ru\\
$^{3}$ \quad Matej Bel University, SK-97401 Banska Bystrica, Slovakia; \\
\phantom{$^3$} \quad Laboratory of Theoretical Physics, Joint Institute for Nuclear Research, RU-141980 Dubna, Russia; evgeni.kolomeitsev@umb.sk
}
\corres{Correspondence: Evgeni Kolomeitsev, evgeni.kolomeitsev@umb.sk}

\abstract{
The strange baryon production in Bi+Bi collisions at $\sqrt{s_{NN}}=9.0$\,GeV is studied using the PHSD transport model. Hyperon and anti-hyperon yields, transverse momentum spectra, and rapidity spectra are calculated, their centrality dependence and the effect of rapidity and transverse momentum cuts are studied. The rapidity distributions for $\ALambda$, $\Xi$, $\AXi$ baryons are found to be systematically narrower than for $\Lambda$s. The $p_T$ slope parameters for anti-hyperons vary more with centrality than those for hyperons. Restricting the accepted rapidity range to $|y|<1$ increases the slope parameters by 13--30\,MeV depending on the centrality class and the hyperon mass. Hydrodynamic velocity and vorticity fields are calculated and the formation of two oppositely rotating vortex rings moving in opposite directions along the collision axis is found. The hyperon spin polarization induced by the medium vorticity within the thermodynamic approach is calculated and the dependence of the polarization on the transverse momentum and rapidity cuts and on the centrality selection is analyzed. The cuts have stronger effect on the polarization of $\Lambda$ and $\Xi$ hyperons than on the corresponding anti-hyperons. The polarization signal is maximal for the centrality class 60-70\%. We show that for the considered hyperon polarization mechanism
the structure of the vorticity field makes an imprint on the polarization signal as a function of the azimuthal angle in the transverse momentum plane, $\phi_H$, $\cos\phi_H=p_x/p_T$. For particles with positive longitudinal momentum, $p_z>0$, the polarization increases with $\cos\phi_H$, while for particles with $p_z<0$ it decreases.}

\keyword{heavy-ion collisions, hydrodynamics, vorticity, hyperon polarization, vortex rings, dynamical freeze-out, NICA, PHSD, MPD}

\def\vec#1{\boldsymbol{ #1}}

\def\om{\omega}
\def\rmd{{\rm d}}

\def\ALambda{\overline{\Lambda}}
\def\AXi{\overline{\Xi}}
\def\AOmega{\overline{\Omega}}

\DeclareMathOperator{\rot}{rot}

\newcommand{\lsim}{\stackrel{\scriptstyle <}{\phantom{}_{\sim}}}

\begin{document}

\section{Introduction}\label{sec:intro}

The global spin polarization of hyperons has been measured in heavy-ion collisions (HICs) in the broad energy range from several TeV at LHC~\cite{Acharya} to a few GeV at SIS~\cite{Yassine:2022}. The polarization signal increases with decreasing energy, reaching a maximum at center-of-mass energies around $\sqrt{s_{NN}}=3$\,GeV. A remarkable feature of the hyperon polarization is that $\overline{\Lambda}$s are more strongly polarized than $\Lambda$s~\cite{Adamczyk-Nature}. This is in striking contrast to hyperon polarization in proton-proton and proton-nucleus collisions where the anti-hyperons do not gain any polarization~\cite{Felix-L-pol}. Therefore, another mechanism is operative in HICs~\cite{Liang-Wang-PRL94,Ladygin2010}. Several such mechanisms have been proposed in the literature, see review~\cite{Becattini-Lisa-2020}. The statistical approach developed in~\cite{Becattini-Tinti2010,Becattini-Chandra2013,Fang-Pang-Wang2016,Becattini-Karpenko-Lisa2017} suggests that the local spin polarization of fermions arises due to the local vorticity gained by the fireball medium due to the initial angular momentum of the colliding nuclei.
Implemented in hydrodynamic~\cite{Karpenko-Becattini2017,Xie-Wang-Csernai2017,Ivanov-PRC100,Ivanov-PRC102,Ivanov-PRC103,Ivanov-PRC105} and transport models~\cite{Li-Pang-Wang-Xia-PRC96,Sun-Ko-PRC96,KTV-PRC97,Wei-Deng-Huang-PRC99,Shi-Li-Liao-PLB788,Vitiuk-BZ2020} this mechanism
allows to reproduce the measured $\Lambda$ polarization except for the low energy HADES data, see Fig.~3 in~\cite{Yassine:2022}.
On the other hand, the splitting between $\overline{\Lambda}$ and $\Lambda$ polarization signals cannot be reproduced in most works unless a special mechanism distinguishing particles from anti-particles is introduced~\cite{Csernai-Kapusta-Welle-PRC99,Ivanov-PRC105,Ivanov-PRC102-AVE}.
Within the transport approach, the splitting was reproduced in the UrQMD model~\cite{Vitiuk-BZ2020} and in the PHSD model~\cite{Voronyuk:2023vyu}.
The origin of the splitting is discussed in detail in Ref.~\cite{Tsegelnik:2023isj}.

The Multi-Purpose Detector (MPD) is the flagship heavy-ion experiment to be built at the Nuclotron-based Ion
Collider fAcility (NICA) at JINR (Dubna) in the center-of-mass energy range $\sqrt{s_{NN}}=4$--11\,GeV~\cite{MPD-EPJA58}.
It is well suited for the identification of strange particles~\cite{Drnoyan:2022lfw}. Recently, it has been shown that the MPD detector system is able to reconstruct the global $\Lambda$ polarization~\cite{Nazarova:2024jic}.

In this paper, we consider the strange particle production and the formation of the polarization signal in the collision Bi$+$Bi at $\sqrt{s_{NN}}=9.0$\,GeV.
The potential of the MPD for studying this colliding system was previously outlined in Ref.~\cite{Mudrokh-2024}.
We demonstrate that the vortical mechanism of the hyperon polarization makes a particular impact on the dependence of hyperon polarization on its momentum. We use the Parton-Hadron-String Dynamics (PHSD) transport model~\cite{PHSD,PHSD-contin}, which was extended~\cite{helicity:2022} for the determination of the vorticity field in the fluidized hot and dense subsystem of interacting particles (participants). As shown in Refs.~\cite{Tsegelnik:2023isj,Voronyuk:2023vyu}, it is possible to reproduce the measured global $\Lambda$ polarization in collisions at $\sqrt{s_{NN}}= 7.7$ and 11.5\,GeV and the global $\overline{\Lambda}$ polarization at 11.5\,GeV.

In Section~\ref{sec:hypprod} we define a criterium for the selection of centrality classes and present our results of hyperon yields and their rapidity and transverse momentum distributions. In Section~\ref{sec:polar} we calculate the hyperon polarization and examine the dependence of the polarization signal for various hyperon species on the transverse momentum and rapidity cuts. Furthermore, we examine the centrality dependence of the hyperon polarization signals.
In Section~\ref{sec:correlat} we investigate the correlations between the hyperon polarization signal and its momentum. Conclusions are formulated in Section~\ref{sec:concl}.

\section{Hyperon multiplicities and spectra}\label{sec:hypprod}

Since in HICs we deal with finite nuclear systems with a non-homogenous density distribution the results of any measured observables depend on the initial geometry. The initial energy/entropy content of the nucleus overlap zone crucially determines final particle multiplicities and momenta distribution and/or energy, measured in the forward rapidity region, which is sensitive to spectator fragments. Therefore, to gain a better understanding of the processes in HICs and constrain model parameters, one is interested in selection of the initial configuration of colliding nuclei casted, e.g., in terms of the impact parameter $b$, or mediate quantities such as the number of nucleon collisions $N_{\rm coll}$ and the number of participating nucleons $N_{\rm part}$.

In experimental studies, collisions are grouped into event (centrality) classes when the most central class includes the events with the highest multiplicity of secondary particles (smallest forward energy), which correspond to small values of the impact parameter.

For the centrality determination, one constructs the event distribution of charged particles, $\frac{\rmd N_{\rm ev}}{\rmd N_{\rm ch}}$,
using some microscopic model of a collision or the Glauber Monte Carlo method~\cite{Miller2007-Glauber}, and, then, defines the centrality classes as a fraction of the total integral
\begin{align}
C=\frac{1}{N_{\rm ev}} \intop_{0}^{N_{\rm ch}} \frac{\rmd N_{\rm ev}}{\rmd N_{\rm ch}} \rmd N_{\rm ch}\,.
\label{Central-def}
\end{align}
Finally, the geometric parameters such as the impact parameter, $b$, the number of participants, $N_{\rm part}$, and the number of binary nucleon-nucleon collisions can be estimated for each centrality class. The standard experimental procedure is described, for instance, in works~\cite{PhysRevC.86.054908, PhysRevC.88.044909, Zherebtsova:epj182, Klochkov:2017oxr}. In Refs.~\cite{Das-Giacalone-PRC97,Rogly-Giacalone-PR98} another method for the centrality class selection was proposed, which requires only a conjecture of the conditional probability distribution of $N_{\rm ch}$ for the fixed impact parameter $b$. The gamma-distribution was suggested in Ref.\cite{Rogly-Giacalone-PR98} to be used for the probability $P(N_{\rm ch }| b)$. This approach is also applied for the centrality determination in the BM$@$N experiment at NICA~\cite{Segal2023-YaF,Parfenov-Particle21}. As pointed out in Ref.\cite{Segal2024}, the centrality determination could be influenced by the detector acceptance (for most central collisions) and by the spectator fragmentation.

\begin{figure}
	\centering
	\includegraphics[width=6cm]{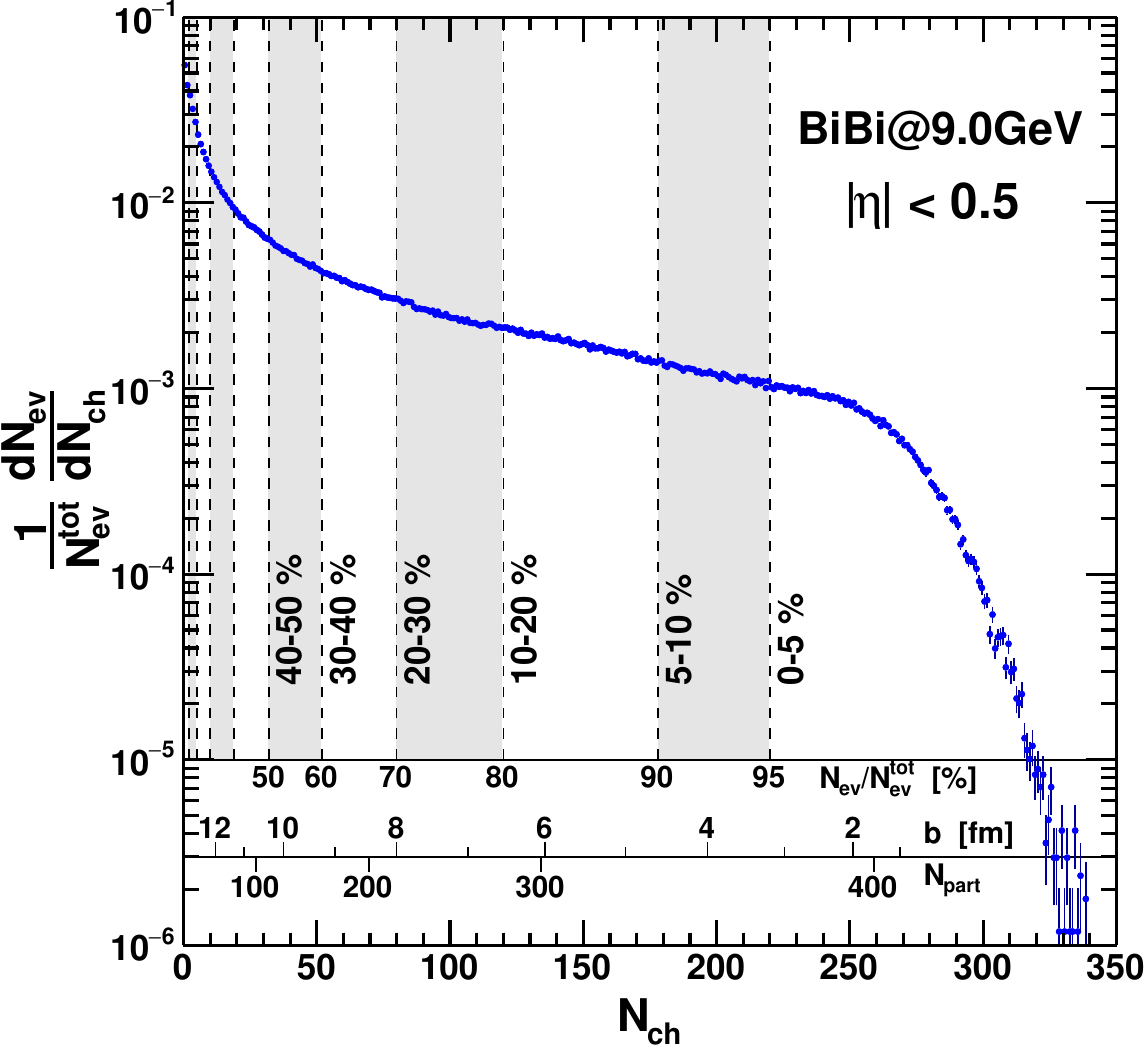}
	\caption{The multiplicity histogram for charged particles $N_{\rm ch}$ in the middle pseudorapidity region $|\eta| < 0.5$ in the Bi+Bi collisions at $\sqrt{s_{NN}}=9.0\,$GeV simulated within the PHSD model. The centrality classes shown by vertical lines are determined using a part of the total integral of the histogram, see Eq,~(\ref{Central-def}). Two additional internal scales are displayed on the plot, indicating the number of participants, $N_{\rm part}$, and the impact parameter, $b$, of the collision.}
	\label{fig:centrality}
\end{figure}

Using the PHSD model, we generated about $N_{\rm ev}^{\rm tot}\simeq 2\times 10^6$ events. The distribution of the number of charged particles, $N_{\rm ch}$,
with the pseudorapidity range $|\eta| < 0.5$ is shown in Fig.~\ref{fig:centrality}. We define the centrality classes using the procedure similar to the traditional experimental one (\ref{Central-def}). The division in centrality classes is presented in Fig.~\ref{fig:centrality}. We also show the corresponding number of participants $N_{\rm part}$ and impact parameter for certain $N_{\rm ch}$.

The PHSD transport model was developed to reproduce the observed strange particle multiplicities in the broad range of collision energies~\cite{Cassing-PR308-HSD,PHSD,Cassing-Palmese-PRC93,Palmese:2016rtq}. In-medium modifications of kaon properties were needed to understand the strangeness production at SIS energies, $\sqrt{s_{NN}}\simeq 2\mbox{--}3$\,GeV. For higher collision energies (in the SPS and RHIC energy range) another mechanism is implemented in the description of initial hard processes taken as in the FRITIOF Lund model~\cite{NILSSONALMQVIST1987387,Andersson:1992iq}, which utilizes the Schwinger mechanism for the quark-anti-quark pair production in string decays. References~\cite{Cassing-Palmese-PRC93,Palmese:2016rtq} suggested in-medium modifications of the quark masses due to a decrease of the quark condensate magnitude when the string breaking occurs in the dense and hot medium created in collisions. This leads to an increase of the probability of the strangeness production. Within the PHSD code, the values of quark condensate are calculated in each cell on each time step. These modifications of the string-breaking processes lead to significant improvements in the description of strange particle production at AGS and SPS energies~\cite{Cassing-Palmese-PRC93,Palmese:2016rtq}.

\begin{figure}
	\centering
\includegraphics[width=14cm]{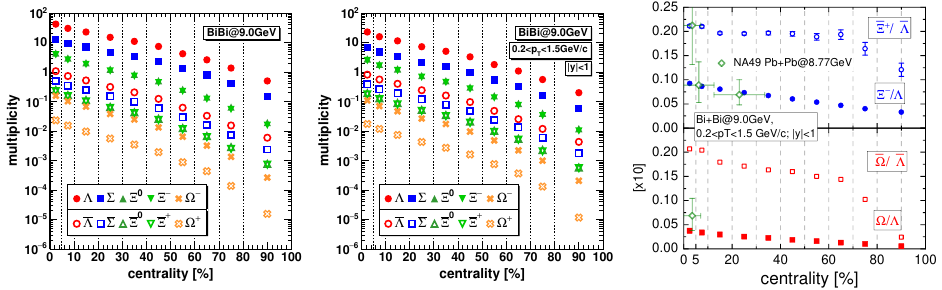}
	\caption{Hyperon multiplicities for various centrality classes in Bi+Bi collisions at $\sqrt{s_{NN}}=9.0\,$GeV without any cuts (left panel) and in the midrapidity region $|y|<1$ with transverse momentum cut $0.2<p_{T}<1.5\,$GeV$/c$ (middle panel). The vertical dashed lines show the widths of the centrality bins. The minimum bias multiplicity almost coincides with one from the centrality class 30--$40\%$. Statistical error bars are of a size of a symbol.
Right panel shows the centrality dependence of the hyperon multiplicities ratios. Open diamonds correspond to the experimental ratios of hyperon multiplicities at the mid-rapidity measured by the NA49 collaboration~\cite{Alt-PRC78,Alt-PRL94} in Pb+Pb collisions at $\sqrt{s_{NN}}=8.77$\,GeV ($40\,A$GeV beam energy).}
\label{fig:multiplicity}
\end{figure}

In Fig.~\ref{fig:multiplicity} we present multiplicities of hyperons and anti-hyperons (i.e. averaged number of particles per event, $N_{H}/N_{\rm ev}^{\rm tot}$) with and without experimental $p_T$ and rapidity cuts for various centrality classes. Calculations show that the results for the minimum bias collisions correspond to the centrality class 30--40\%. The multiplicity drops exponentially with a centrality increase, i.e. for larger impact parameters and more peripheral collisions. So, the minimum bias multiplicities are on average 3.3 times smaller than for the most central collisions of the 0-10\% centrality class. For the most central collisions with the 0-5\% centrality class, we have the following relations among yields of hyperons with the various strangeness content
\begin{align}
&M_\Lambda : M_\Xi : M_\Omega = 1:(9.10\pm 0.0  2)\times 10^{-2}:(3.71\pm 0.04)\times 10^{-3} \,,
\nonumber\\
&M_{\overline{\Lambda}} : M_{\overline{\Xi}} : M_{\overline{\Omega}} = 1:(2.09\pm0.02)\times 10^{-1}:(2.16\pm 0.06)\times 10^{-2} \,.
\label{relat-yields}
\end{align}
This hierarchy of yields can be compared with the results of the NA49 experiment on central (0--7\%) Pb+Pb collision at $\sqrt{s_{NN}}=8.77$\,GeV
reported in Refs.~\cite{Alt-PRC78,Alt-PRL94}
\begin{align}
M_\Lambda : M_\Xi : M_\Omega &= 1:(7\pm 1)\times 10^{-2}:(3\pm2)\times   10^{-3}\,,
\nonumber\\
M_{\overline{\Lambda}} : M_{\overline{\Xi}} &= 1:(2\pm 0.5)\times 10^{-1}\,.
\label{exp-yields}
\end{align}
There is a good overall agreement between our calculations (\ref{relat-yields}) and experimental finding for collisions of nuclei with similar $A$ and similar collision energies.

The imposed rapidity and $p_T$ cuts reduce the number of $\Lambda$s and $\Sigma$s by a factor of 0.6, $\Xi$ hyperons by a factor of 0.7, and $\Omega$ hyperons by a factor of 0.8. The anti-hyperons are influenced by the cuts even weaker: mutiplicities of $\overline{\Lambda}$, $\overline{\Sigma}$, $\overline{\Xi}$, and $\overline{\Omega}$ are reduced by a factor of 0.9.

The relative enhancement of multi-strange particle production in central heavy-ion collisions with respect to peripheral ones has been suggested as a signature for the transient existence of a QGP phase~\cite{Koch:1986ud}. Such a weak enhancement of the $\Xi^-/\Lambda$ ratio experimentally observed in collisions at the AGS energies was analyzed in Ref.~\cite{Zeeb:2003wv} within the UrQMD transport model. For the top SPS energy, the more pronounced centrality dependence of $\Xi^-/\Lambda$, $\Omega/\Lambda$, $\overline{\Xi}^+/\ALambda$, and $\AOmega/\ALambda$ ratios was described in Ref.~\cite{Arakelyan-PRD105} in the framework of the QGSM model~\cite{Amelin:1989wa,Amelin:1990js,Amelin:1989ve}. In the right panel of Fig.~\ref{fig:multiplicity} we show the ratios of hyperon and anti-hyperon multiplicities. For comparison, we depict in Fig.~\ref{fig:multiplicity} the ratios of experimental multiplicities measured by the NA49 collaboration in Pb+Pb collisions at $\sqrt{s_{NN}}=8.77$\,GeV~\cite{Alt-PRC78,Alt-PRL94}. We used the mid-rapidity multiplicities, for which there are some information on the centrality dependence~\cite{Alt-PRC78}. We see in Fig.~\ref{fig:multiplicity}, first, that the ratios of anti-particles is larger than the particle ratios, i.e. $\frac{\AXi^+}{\ALambda} > \frac{\Xi^-}{\Lambda}$ and $\frac{\AOmega}{\ALambda} > \frac{\Omega}{\Lambda}$. Second observation is that the anti-hyperon ratios increase very rapidly from the most peripheral to the semi-central collisions for $60\%<C<100\%$. With a further decrease of $C$ below 60\%, all ratios grow slowly.

\begin{figure}
	\centering
	\includegraphics[width=10cm]{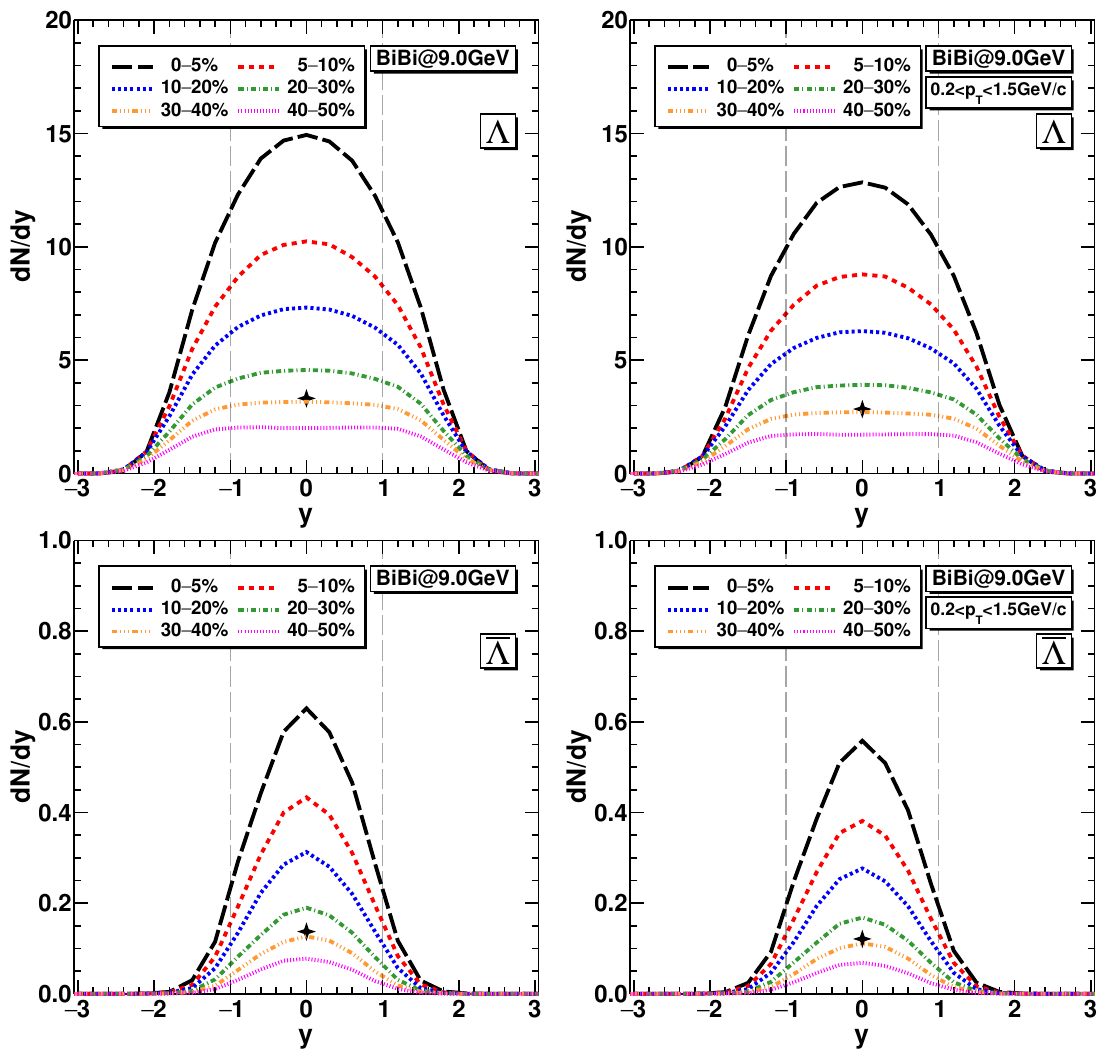}
	\caption{Rapidity distribution of $\Lambda$ hyperons (upper row) and $\ALambda$ hyperons (lower row) without momentum cuts (left) and with the transverse momentum cut $0.2<p_{T}<1.5\,$GeV$/c$ (right). The minimum bias value $dN/dy|_{y=0}$ is shown by a star. The statistical uncertainty of the calculations is comparable with the line width.}
\label{fig:lambda:y-dist}
\end{figure}

To clarify which of the cuts is responsible for the yield suppression to a larger expansion, we consider rapidity and $p_T$ distributions of hyperons.
Figure~\ref{fig:lambda:y-dist} presents the rapidity distributions of $\Lambda$ and $\ALambda$ with and without $p_T$ cuts.
Comparing plots in the left and right columns we see that the $p_T$ cuts lead only to a moderate reduction of particle yields: by a factor of 0.86 for $\Lambda$ particles and by a factor of 0.92 for $\ALambda$. The strongest effect is caused by the rapidity cut shown in Figure~\ref{fig:lambda:y-dist} by vertical lines. We also observe that the influence of this cut is stronger for $\Lambda$ hyperons than for $\ALambda$, since the former one has a broader rapidity distribution. For a more quantitative comparison of the rapidity distributions we parameterize them by two Gaussian functions with a width $\sigma$ shifted by $\Delta y$ around the midrapidity,
\begin{align}
\frac{\rmd N}{\rmd y}=\frac{A}{2\sqrt{2\pi\sigma^2}}\Big(\exp\Big[-\frac{(y-\Delta y)^2}{2\sigma^2}\Big]+\exp\Big[-\frac{(y+\Delta y)^2}{2\sigma^2}\Big]\Big)\,.
\label{dNdy-fit-fun}
\end{align}
Using this distribution we fit the $\Lambda$ and $\ALambda$ rapidity distributions shown in Fig.~\ref{fig:lambda:y-dist} for the centrality classes 0--5\% and 30--40\% in the rapidity range $-2\le y\le 2$ and obtain for $\Lambda$ hyperons
\begin{align}
\mbox{centrality class 0--5\%}:\quad & \Delta y_\Lambda = 0.72(1)\,,\quad \sigma_\Lambda=0.74(2)\,,\quad A_\Lambda=44\pm 0.5\,,
\nonumber\\
\mbox{centrality class 30--40\%}:\quad& \Delta y_\Lambda = 0.91(1)\,,\quad \sigma_\Lambda=0.82(3)\,,\quad A_\Lambda=12\pm 0.2\,,
\label{y-dist-fit-L}
\end{align}
and for $\ALambda$ hyperons
\begin{align}
\mbox{centrality class 0--5\%}:\quad & \Delta y_{\ALambda} = 0.41(1)\,,\quad \sigma_{\ALambda}=0.49(1)\,,\quad A_{\ALambda}=1.08\pm 0.01\,,
\nonumber\\
\mbox{centrality class 30--40\%}:\quad & \Delta y_{\ALambda} = 0.39(1)\,,\quad \sigma_{\ALambda}=0.46(1)\,,\quad A_{\ALambda}=0.206\pm 0.001\,.
\label{y-dist-fit-AL}
\end{align}
We see that the rapidity distributions of $\ALambda$ are significantly narrower than those for $\Lambda$, i.e  $\Delta y_{\ALambda} < \Delta y_\Lambda$ and
$\sigma_{\ALambda}< \sigma_{\Lambda}$. Another interesting difference between hyperons and anti-hyperons is that the $\Lambda$ $y$-distributions for 30--40\% centrality is broader than that for 0--5\% centrality: $\Delta y_\Lambda[\mbox{0--5\%}]< \Delta y_\Lambda[\mbox{30--40\%}]$ and
$\sigma_\Lambda[\mbox{0--5\%}] < \sigma_\Lambda[\mbox{30--40\%}]$, whereas for $\ALambda$ we observe the opposite dependence:
$\Delta y_{\ALambda}[\mbox{0--5\%}] > \Delta y_{\ALambda}[\mbox{30--40\%}]$ and $\sigma_{\ALambda}[\mbox{0--5\%}] > \sigma_{\ALambda}[\mbox{30--40\%}]$.

\begin{figure}
	\centering
	\includegraphics[width=10cm]{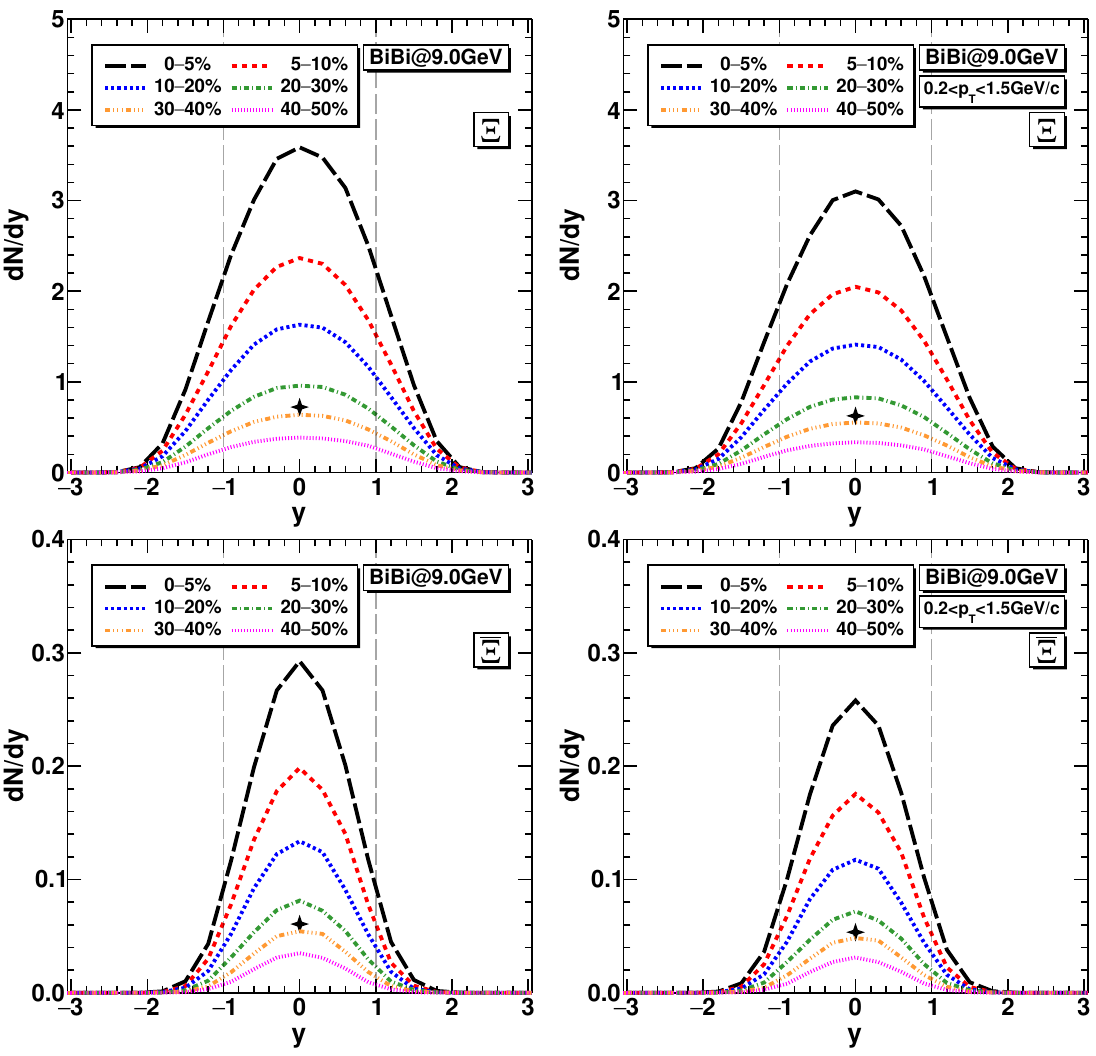}
	\caption{The same as in Fig.~\ref{fig:lambda:y-dist} but for $\Xi$ and $\AXi$ hyperons.}
	\label{fig:xi:y-dist}
\end{figure}

In Fig.~\ref{fig:xi:y-dist} we show the rapidity distributions of $\Xi$ and $\AXi$ hyperons. Fitting these distributions with Eq.~(\ref{dNdy-fit-fun})
we find for $\Xi$ hyperons
\begin{align}
\mbox{centrality class 0--5\%}:\quad & \Delta y_\Xi = 0.55(1)\,,\quad \sigma_\Xi=0.65(2)\,,\quad A_\Xi=3.91\pm 0.04,
\nonumber\\
\mbox{centrality class 30--40\%}:\quad & \Delta y_\Xi = 0.59(1)\,,\quad \sigma_\Xi=0.67(2)\,,\quad A_\Xi=0.736\pm 0.006,
\label{y-dist-fit-X}
\end{align}
and for $\AXi$ hyperons
\begin{align}
\mbox{centrality class 0--5\%}:\quad & \Delta y_{\AXi} = 0.38(1)\,,\quad \sigma_{\AXi}=0.47(1)\,,\quad A_{\AXi}=0.226\pm 0.001,
\nonumber\\
\mbox{centrality class 30--40\%}:\quad & \Delta y_{\AXi} = 0.37(1)\,,\quad \sigma_{\AXi}=0.42(2)\,,\quad A_{\AXi}=0.040\pm 0.001.
\label{y-dist-fit-AX}
\end{align}
Hence, the $\Xi$ rapidity distributions are narrower than those for $\Lambda$s, whereas the $\AXi$ distributions are similar to those for $\ALambda$.
Remarkably, both $\Xi$ and $\AXi$ distributions depend very weakly on centrality.

\begin{figure}
	\centering
\includegraphics[width=10cm]{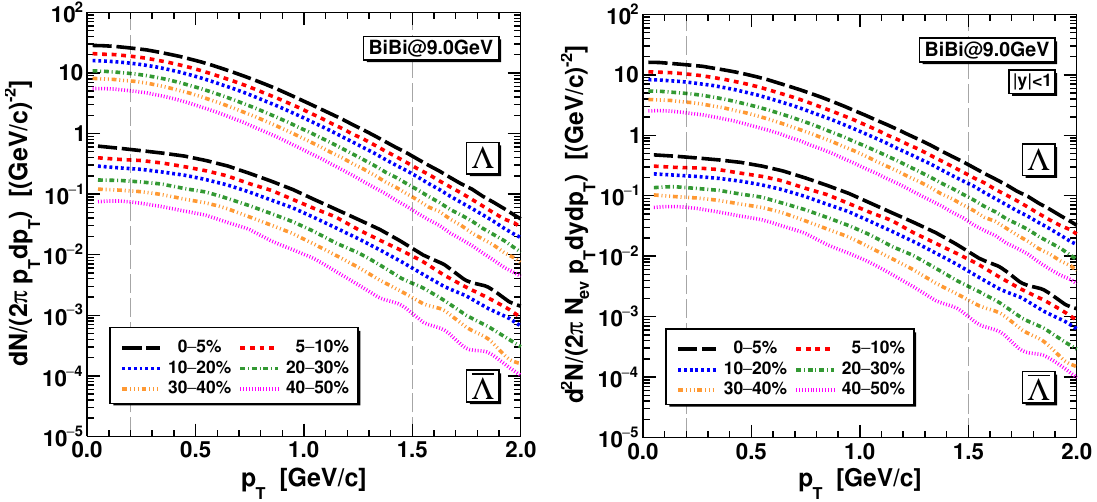}
	\caption{Transverse momentum distributions of $\Lambda$ and $\ALambda$ hyperons for different centrality classes without momentum cuts (left panel) and in the rapidity range $|y|<1$ (right panel).}
\label{fig:lambda:pt-dist}
\end{figure}

\begin{figure}
	\centering
\includegraphics[width=10cm]{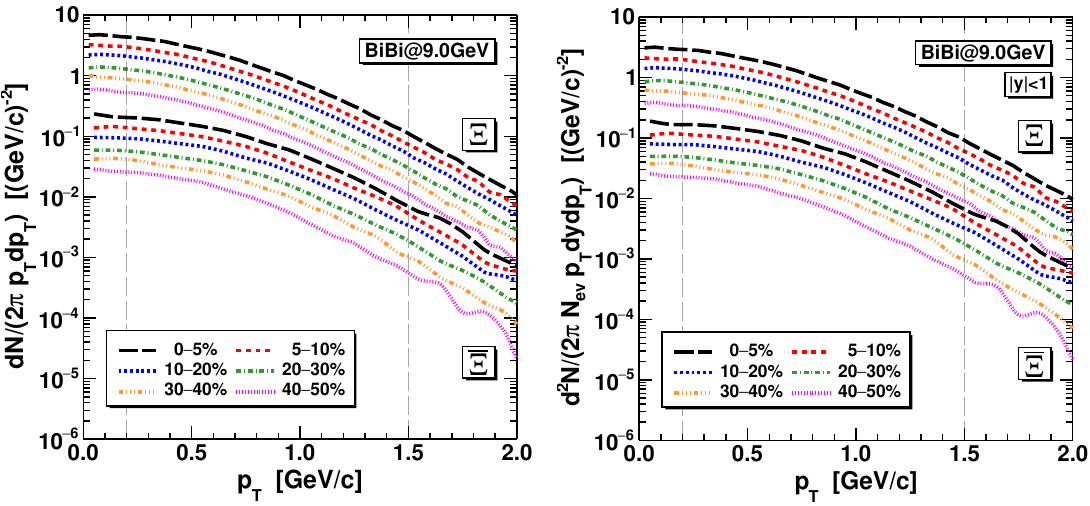}
	\caption{The same as in Fig.~\ref{fig:lambda:pt-dist} but for $\Xi$ and $\AXi$ hyperons.}
	\label{fig:xi:pt-dist}
\end{figure}

In Ref.~\cite{Tsegelnik:2023isj} we showed that the PHSD code can reproduce the $p_T$ spectra of $\Lambda$s measured by the STAR collaboration in Au+Au collisions~\cite{Adam:2019koz} at $\sqrt{s_{NN}}=7.7$\,GeV. In Ref.~\cite{Voronyuk:2023vyu} the excitation functions of the hyperon multiplicities are confronted with available data for AGS, SPS, and RHIC energies. Further comparisons of the PHSD code with the experimental data can be found in~\cite{Cassing-Palmese-PRC93,Palmese:2016rtq}. For the Bi+Bi collisions at 9.0\,GeV we consider here, the transverse momentum spectra of $\Lambda$ and $\ALambda$ hyperons are depicted in Fig.~\ref{fig:lambda:pt-dist}, and the spectra of $\Xi$ and $\AXi$ hyperons in Fig.~\ref{fig:xi:pt-dist}. For momenta below 1.5\,GeV the statistical uncertainty is of the size of the line width but it increases for higher momenta as follows from the line oscillations especially for anti-hyperons. The magnitudes of the spectra are reflected in the total hyperon multiplicities discussed above, see Fig.~\ref{fig:multiplicity}. To quantify the slopes of the spectra we fit them in the momentum interval $0.5\le p_T\le 1.5$\,GeV with the blast-wave formula~\cite{Schnedermann-Fireball,Schnedermann:1993ws},
\begin{align}
\frac{\rmd N}{2\pi p_T\rmd p_T}=A m_T K_1\big(m_T/T_{\rm slope}\big)\,,\quad m_T=\sqrt{m_H^2+p_T^2}\,,
\label{pt-fit}
\end{align}
for spectra without the rapidity cut. For spectra with the rapidity cut $|y|\le 1$ we would have to use expression
\begin{align}
\frac{\rmd N}{2\pi p_T\rmd p_T} = A m_T \intop_{0}^{1}\rmd y \cosh y \exp\Big(\frac{m_T}{T_{\rm slope}}\cosh y\Big).
\label{pt-fit-cut}
\end{align}
However, for the typical parameters $T_{\rm slope}\sim$ 150--200\,MeV the functions (\ref{pt-fit}) and (\ref{pt-fit-cut}) differs by less than 1\%, so we can use Eq.~(\ref{pt-fit}). We should emphasize that the simple parameterization above can be used only for rather narrow momentum intervals. For a broader interval including low and high $p_T$ parts, combinations of several distributions, like a double-temperature Boltzmann distribution together with a Tsallis-Pareto distribution, should be applied.

\begin{figure}
	\centering
\includegraphics[width=10cm]{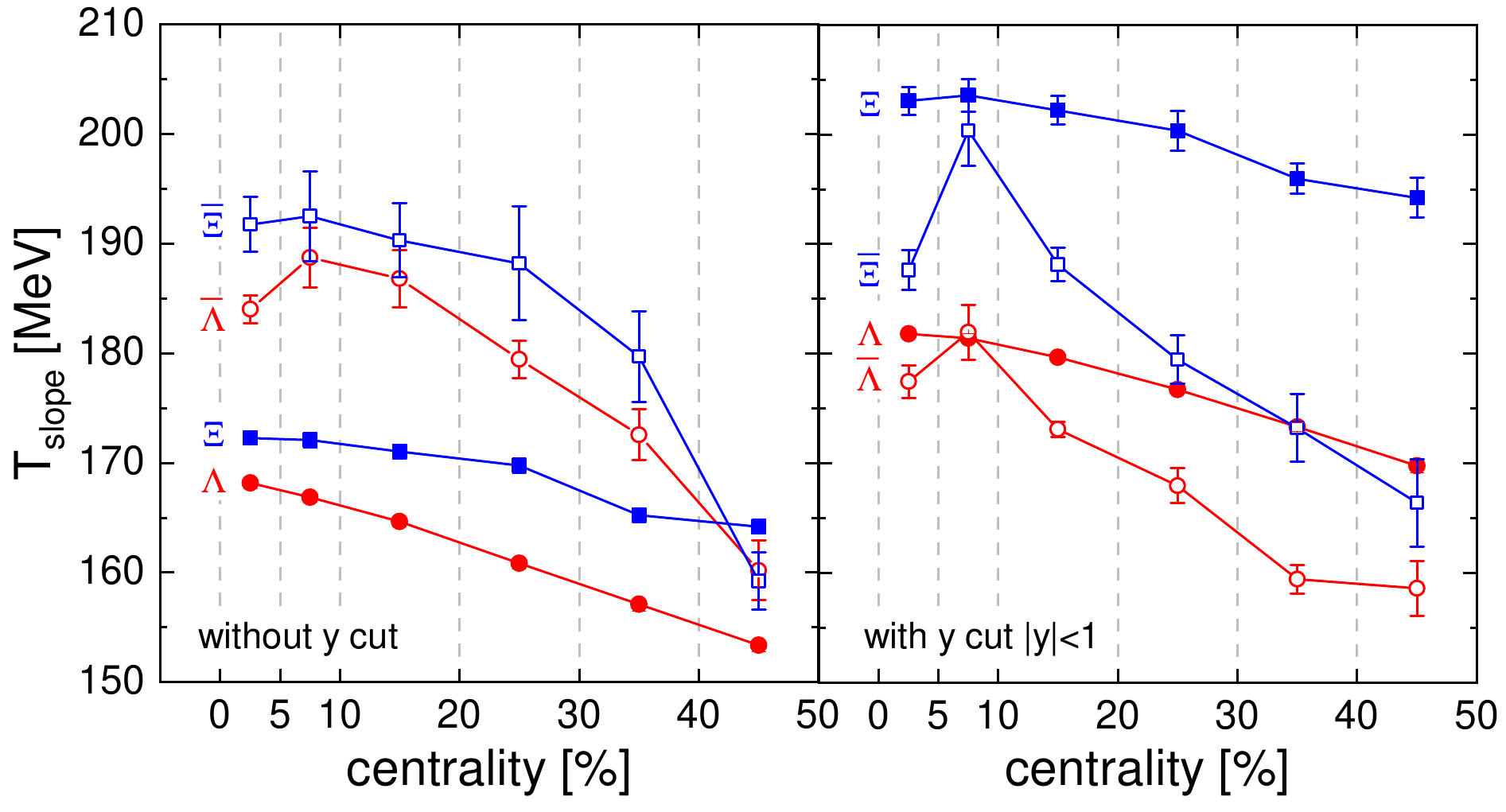}
	\caption{Slope parameters of transverse momentum distributions of $\Lambda$, $\Xi$, $\ALambda$, and $\AXi$ hyperons within the momentum interval $0.5\le p_T\le 1.5$\,GeV  for different centrality bins without any cuts (left panel) and for the rapidity cut $|y|<1$ (right panel). If error bars are not visible they are smaller than the symbol size.}
	\label{fig:Tslope}
\end{figure}

The results of fits are presented in Fig.~\ref{fig:Tslope}. The slope parameters for $\Lambda$ and $\Xi$ hyperons show a weak dependence on the centrality monotonously decreasing by $\sim 15$\,MeV for $\Lambda$s and  $\sim 8$\,MeV for $\Xi$s from the most central collisions to the 40-50\% centrality class. Anti-hyperons show much stronger variations with the centrality increase by about 30\,MeV from central to semi-peripheral collisions. Also, the variation is not monotonous and the slope parameter increases, first, from the 0-5\% centrality bin to the 5-10\% bin. Comparing the left and right planes in Fig.~\ref{fig:Tslope} we see that restricting rapidities to $|y|<1$ leads to an increase of the slope parameters by 13--17\,MeV for $\Lambda$s and by 30\,MeV for $\Xi$s. The centrality dependence of the slope parameter is similar to those without the rapidity cut. For anti-hyperons, the restriction of the rapidity range produces a smaller increase of the $T_{\rm slope}$ but enhances the non-monotonous variation of the slope parameter with centrality.

Both the slope of $p_T$ spectra and the width of rapidity spectra suggest that observed anti-hyperons decouple from the fireball mainly at earlier stages of its evolutions than hyperons. This agrees with the analysis of Au+Au collision~\cite{Tsegelnik:2023isj,Voronyuk:2023vyu} where we argued that this explains the larger magnitude of the global spin polarization of anti-hyperons.

\section{Vorticity field and spin polarization}\label{sec:polar}

For the calculations of the hyperon spin polarization we use the thermodynamic approach~\cite{Becattini-Chandra2013,Becattini:2015ska}, where the local spin polarization of a particle with mass $m$ and spin $s$ is induced by the local thermal vorticity:
\begin{equation}\label{eq:becattini:thermal-vorticity}
	\varpi_{\mu\nu} = \frac{1}{2} \Big[\partial_{\nu} \Big(\frac{u_\mu}{T}\Big) - \partial_{\mu} \Big(\frac{u_\nu}{T}\Big)\Big]
=\frac{\omega_{\mu\nu}}{2T} - \frac{1}{2T}\Big[u_\mu\partial_\nu \ln T - u_\nu \partial_\mu\ln T\Big],
\end{equation}
where $u_\nu$ is the hydrodynamic 4-velocity of a fluid element and $T$ is its temperature.
In the second equation here we separated the kinematic vorticity $\omega_{\mu \nu} = (\partial_{\nu} u_{\mu} - \partial_{\mu} u_{\nu})$. The latter tensor provides a natural relativistic generalization for the non-relativistic vorticity $\vec{\om}=\rot \vec{u}$:
$\om_{\mu\nu}=-\epsilon_{\mu\nu\rho\sigma}u^\rho \bar{\om}^\sigma$\,.
The components of the relativistic vorticity vector are $\bar{\omega}^\mu = \gamma^{2} \big((\vec{u}\vec{\om}),\vec{\omega} + [\vec{u} \times \partial_{t} \vec{u}]\big)$.
Then, hyperons with spin $s$, momentum $p_\mu$ and mass $m_H$  contained in this fluid element will have average spin described by the 4-vector
\begin{equation}\label{eq:becattini:S-def}
	S^{\mu}(x,p)=-\frac{s\, (s+1)}{6\, m_H}(1- n(x,p))\varepsilon^{\mu\nu\lambda\delta}\varpi_{\nu\lambda}p_\delta,
\end{equation}
to the leading order in the vorticity $\varpi_{\mu\nu}$. Here $n(x,p) $ is the hyperon distribution function.
Applying the relation (\ref{eq:becattini:S-def}) we do not assume that the system has reached global thermal equilibrium. We determine local thermodynamic parameters of the medium, like temperature and density performing ensemble averaging over many collision events. Particularly, to determine the fluid velocity, $\vec{u}$, and the energy density, $\varepsilon$, we calculate averaged energy-momentum tensor for the particle distribution generated by the PHSD code, and find its eigenvectors $\vec{u}$ and eigenvalues $\varepsilon$ at any moment of the system evolution. The applied procedure is described in detail in Ref.~\cite{helicity:2022}. We have to emphasize that we do not solve hydrodynamic equations as, for instance, in the hybrid approaches like SMASH and UrQMD, and the evolution of the velocity, density, and density fields follows from the microscopic transport calculations, see also Ref.~\cite{Moreau-PhysRevC.100.014911}. Given local vorticity field $\varpi(x)$ we can estimate the averaged spin polarization of the fermion created at this point, for details see Ref.~\cite{Tsegelnik:2023isj}.

Distinguishing features of our approach are, first, the clear separation of nucleon-spectators and nucleon-participants. Only the latter ones contribute their angular momentum to the system leading to the vorticity creation. Second, only the fluid cells with the energy density $>0.05\,{\rm GeV/fm^3}$ are taken into account in the vorticity calculations. By this we eliminate cells with an on-average small number of particles in every collision event to get rid of the areas far from hydrodynamic regime and exclude thereby high fluctuating gradients due to extremely small densities and temperatures.

The obtained range of temperatures and densities is such that hyperon momentum distributions are essentially non-degenerate and, therefore, we can work in the Boltzmann limit putting $(1- n(x,p)) \approx 1$ in Eq.~(\ref{eq:becattini:S-def}).

\begin{figure}
	\centering
\includegraphics[width=13.5cm]{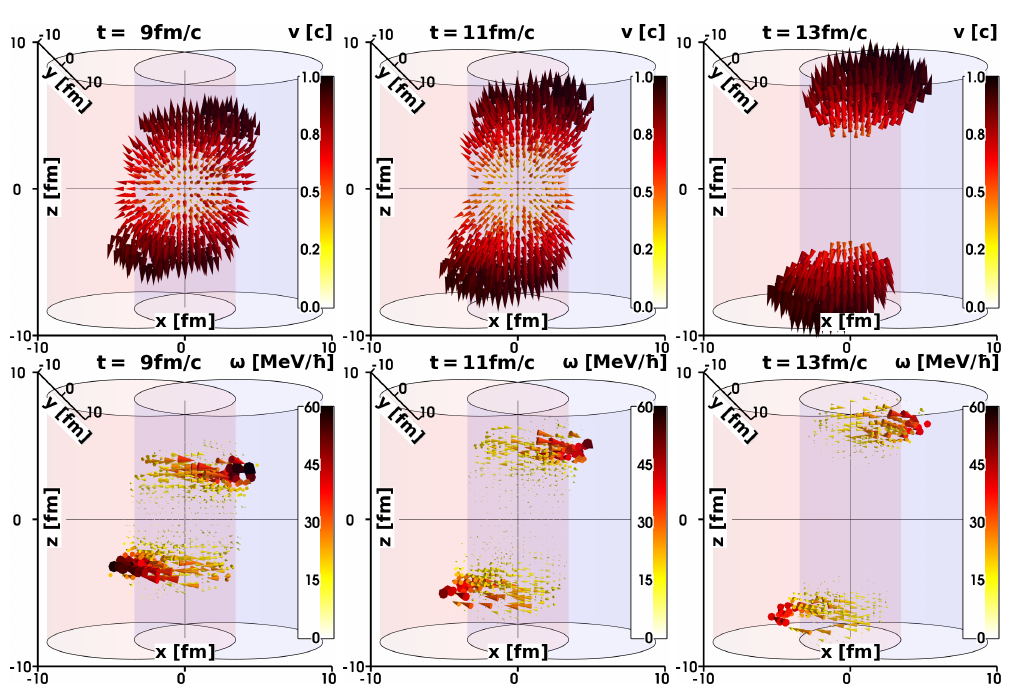}
\caption{Hydrodynamic velocity, $\vec{u}$, (upper row) and corresponding vorticity, $\vec{\omega}=\rot \vec{u}$, (lower row) fields created in Bi+Bi collision at the collision energy $\sqrt{s_{NN}}=9.0$\,GeV and the impact parameter $b=8.5$\,fm at three moments of time. The moment of the full overlap of colliding nuclei corresponds to $\simeq 5.5$\,fm/$c$. Two overlapping cylinders indicate trajectories of colliding nuclei. The $y$ axis points away from the reader so that the $x$, $y$, $z$ axes form the right-handed coordinate system.
}
\label{fig:V-W-fields}
\end{figure}

The velocity and vector fields created in the Bi+Bi collision at $\sqrt{s_{NN}}=9.0$\,GeV are found to be qualitatively very similar to those for Au+Au collision
at $\sqrt{s_{NN}}=7.7$\,GeV which we studied in detail in Refs.~\cite{helicity:2022,Voronyuk:2023vyu}. Therefore, we discuss here only shortly the general structure of these fields which are relevant for the further discussion of the angular dependence of the hyperon polarization. The evolution of the velocity field in Bi+Bi collision at the selected value of the impact parameter $b=8.5$\,fm that corresponds to the centrality class 30--40\%, see Fig.~\ref{fig:centrality}, is shown in Fig.~\ref{fig:V-W-fields} for three time moments. As was shown in Ref.~\cite{helicity:2022} the velocity field consists mainly of the Hubble-like component, which can be written in the cylindrical coordinate system $(r_T,\phi,z)$ as
\begin{align}
\vec{u}_{\rm H} &=\alpha_T\,r_T^{\beta_T} \vec{e}_T + \alpha_{\|}\,z^{\beta_{\|}}\,\vec{e}_z\,,\quad
\label{V-Hubble}
\end{align}
where coefficients $\alpha_T$, $\beta_T$ and $\alpha_{\|}$, $\beta_\|$ do not depend on coordinates (but may change with time) and
$\vec{e}_T =(\cos\phi,\sin\phi,0)$ and $\vec{e}_z=(0,0,1)$ are the unit vectors in the transverse and longitudinal directions.
The evolution of transverse and longitudinal components of the velocity are presented in Figs.~9 and 10 in Ref.~\cite{helicity:2022}.

The Hubble-like component is irrotational, i.e. $\rot \vec{u}_{\rm H}=0$. The strict mathematical Stokes-Helmholtz decomposition of a vector field in irrotational and solenoidal parts is rather computationally involved and is, actually, not necessary for the study of spin polarization.
Some approximate methods of separation of the Hubble-like component we discussed in Ref.~\cite{Hubble}.
The analysis in~\cite{helicity:2022} of the velocity fields generated with the fluidized PHSD model shows that on average the correction terms are numerically much smaller than the Hubble-like term: $|\vec{u}-\vec{u}_{\rm H}|\ll \vec{u}_{\rm H}$. The same we find also for the Bi+Bi collisions.
There are several sources for a small admixture of a vortical component, $\vec{u}_{\om}$, i.e. $\rot \vec{u}_\om =\vec{\om}\neq 0$, which we discuss  qualitatively in order to justify the observed structure of the vorticity field.
A contribution to the vorticity occurs if the coefficient $\alpha_T$ and $\alpha_{\|}$ in (\ref{V-Hubble}) acquire a residual dependence on $z$ and $r_T$, respectively, than $\rot \vec{u}_H=-\big( z\partial_{r_T} \alpha_{\|} -r_T \partial_z \alpha_{T} \big)\vec{e}_\phi$\,. We see it creates a circular vorticity field. Another contribution is due to the term violating the axial symmetry and corresponding to a time and $z$-coordinate shift of the center of the transverse Hublle-like expansion, $(x_0,0)$: $\delta \vec{u}_{\rm asym} = -\alpha_T x_0(z,t)\Big(\cos\phi\vec{e}_T - \sin\phi\vec{e}_\phi\Big)\,,$
where $\vec{e}_\phi$ is the azimuthal unit vector $\vec{e}_\phi=[\vec{e}_z\times \vec{e}_T]=(-\sin\phi,\cos\phi,0)$.
The shift of the center of the transverse expansion is visible in the upper row in Fig.~\ref{fig:V-W-fields}. Such a tilt in the particle emission source with respect to collision axis could be tested, e.g., by the two-pion momentum correlations, see~\cite{LISA20001,Lisa:2000ip}.
Contribution $\delta \vec{u}_{\rm asym}$ to the fluid velocity produces the axially asymmetric vorticity field
$\rot \delta \vec{u}_{\rm asym} = -\alpha_T\partial_z x_0(z,t)\big(\sin\phi\vec{e}_T + \cos\phi \vec{e}_\phi  \big)=-\alpha_T\partial_z x_0(z,t) (1,0,0)$.
Both of the above effects lead to the formation of an asymmetric vortex ring, which has a form of a bublik (an asymmetric donut), see detailed investigation in~\cite{helicity:2022}. The formation of vortex rings was proposed also in Ref.~\cite{Ivanov-rings}. The vorticity field generated by the PHSD model is shown in Fig.~\ref{fig:V-W-fields} and demonstrates clearly two bubliks moving in the opposite direction along the $z$ axis and having the opposite direction of vorticities. The bubliks are slowly expanding in the transverse direction. The details on the structure of the vorticity field and its evolution can be seen in Fig~13 in Ref.~\cite{helicity:2022}. For the Bi+Bi collisions at the energies considered in this paper, the maximum magnitude of the kinematic vorticity over the fireball grows, first, from the time of the full nucleus overlap ($t\sim 5$\,fm/$c$) to $t\sim 7$\,fm/$c$ between $\om\simeq 70$\,MeV/$\hbar$ and 80\,MeV/$\hbar$. Then, it gradually decreases down to $\simeq 40$\,MeV at $t\sim 13$\,fm/$c$.

In Fig.~\ref{fig:V-W-fields} we can note that the maximum of the vorticity magnitude is located at the border of the intersection of colliding nuclei.
This can be understood if we recall the equation for the vorticity dynamics, which follows from the Navier-Stokes equation~\cite{Kundu-FluidDynamics}, see also Ref.~\cite{helicity:2022},
\begin{align}
\frac{\partial \vec{\om}}{\partial t} + [\vec{\nabla} \times [\vec{\om}\times \vec{v}]] =  \nu \Delta \vec{\om} + \frac{1}{\rho^2} [\vec{\nabla} \rho \times \vec{\nabla} p]\,,
\label{Biermann}
\end{align}
where the first term on the right-hand side with the kinematic shear viscosity, $\nu$, is responsible for the decay of the vorticity due to diffusion. The second term is the vorticity source term. This term is called the Biermann battery following Ref.~\cite{Biermann}, where a similar term was considered as a source of the magnetic field in stars. The source term is non-zero when the pressure is nonbarotropic, i.e., depends not only on the particle density but also on the temperature (or entropy) of the system, as it is for the nuclear equation of state. Thus, the vorticity is generated when a density gradient is not collinear with a temperature gradient, $[\nabla \rho\times\nabla p]\propto [\nabla \rho\times\nabla T]\neq 0$. This occurs at the initial stage of the fireball formation when colliding nuclei are interpenetrating at the border of their contact.

\begin{figure}
	\centering
\includegraphics[width=12cm]{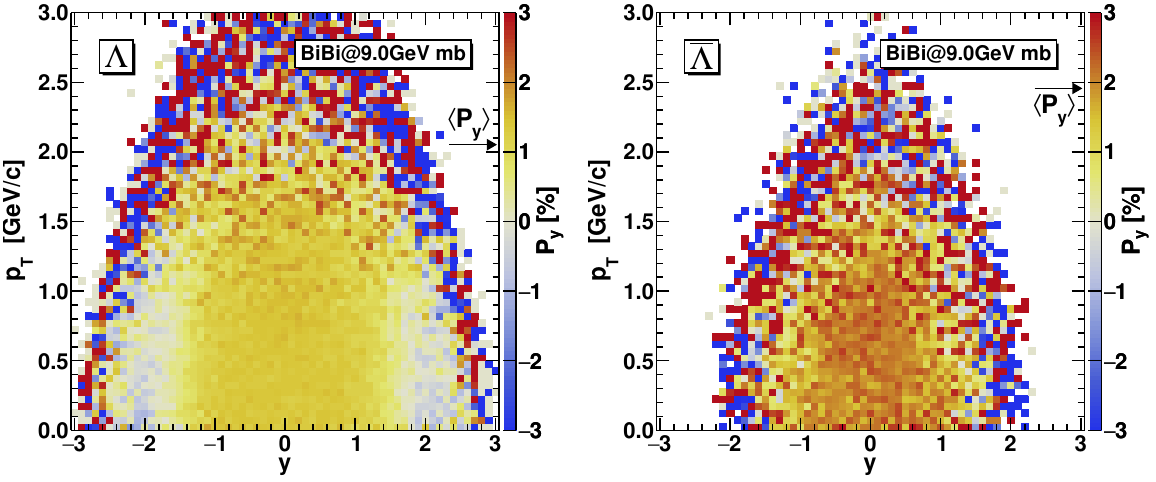}
	\caption{The spin polarization distributions as functions of $p_T$ and $y$ for $\Lambda$ (left panel) and $\ALambda$ (right panel) hyperons for the minimum bias Bi+Bi collisions at $\sqrt{s_{NN}}=9.0\,$GeV. The average polarizations for these distributions are shown by arrows on the scale.
}
\label{fig:pol-map}
\end{figure}

Within our dynamical freeze-out approach, at each time step, if a particle occurs inside the medium as a result of a non-decay process like particle creation or rescattering, then the particle's spin vector is calculated according to Eq.~(\ref{eq:becattini:S-def}). Recall that, under the medium we understand only the fluid elements with the energy density $\varepsilon> 0.05\,{\rm GeV/fm^3}$. The hyperon polarization $\vec{P}=\vec{S^*}/s$ is then determined by the spin vector recalculated in the rest frame of the hyperon:
\begin{equation}\label{eq:becattini:S-boosted-vec}
\vec{S^*} = \vec{S} - \frac{(\vec{S} \vec{p})\, \vec{p}}{E(E+m)},
\end{equation}
where we used $S_0 E=(\vec{S} \vec{p}) $ since $S^\mu p_\mu \equiv 0$. If a particle ends up outside the medium, the particle polarization is set to zero. This means that such particles will not contribute to the accumulated spin of hyperons.

In the PHSD model, strong hyperon decays are dynamically included in the evolution of the system. For strong decays $\Sigma^{*} \rightarrow \Lambda + \pi $ and $\Xi^{*} \rightarrow \Xi + \pi,$ the polarization of the initial hyperon is in part transferred to the daughter hyperon according to the relations derived in Ref.~\cite{Becattini-Karpenko-Lisa2017}, see Section 4 of Ref.~\cite{Voronyuk:2023vyu} for detail. They are based on the consideration that the momentum-independent mean polarization of the daughter fermion is proportional to that of the parent one, and a dynamical matrix independent coefficient can be used, see Eq.~(37) and Table~1 in Ref.~\cite{Becattini-Karpenko-Lisa2017} and appendices there.
Finally, hyperons that survive till the full dissolution of the medium will carry information about the medium and polarization at the time of their last interaction, see Ref.~\cite{Voronyuk:2023vyu} for detail.

Applying this algorithm to $\Lambda$ or $\ALambda$ hyperons and averaging over all generated collision events we can determine the averaged global polarizations of hyperon as the $y$ projection of the vector
\begin{equation}
\langle\vec{P}_{H}\rangle = 2\, \langle\vec{S^*}_{H}\rangle.
\end{equation}
The $y$ axis is always normal to the reaction plane in our simulations. This is primary and not yet a final observable polarization signal, since the weak and electromagnetic decays, $\Xi \rightarrow \Lambda + \pi$ and $\Sigma^0 \rightarrow \Lambda + \gamma$, are not yet taken into account.

In Fig.~\ref{fig:pol-map} we present polarization maps of the transverse momentum $p_T$ and rapidity $y$ for the minimum bias Bi+Bi collisions. One can see a plateau of positive polarization in the $|y|\lsim1.5$ and $p_T\lsim1.5\,$GeV/$c$ area for hyperons and $|y|\lsim1$ and $p_T\lsim1\,$GeV/$c$ for anti-hyperons. Peripheral zones of rapidity and transverse momentum have large fluctuations of polarization in the neighbouring cells. There are very few particles in this cells and the energy density is below the threshold value $0.05\,{\rm GeV/fm^3}$. Therefore, hyperons occurring in these cells acquire the zero polarization. With a further increase of the number of collision events, the contributions from these zones to the total polarization average to zero. By using such maps separated for different centrality classes we can study the influence of experimental cuts on the observable polarization signal.

\begin{figure}
	\centering
\includegraphics[width=10cm]{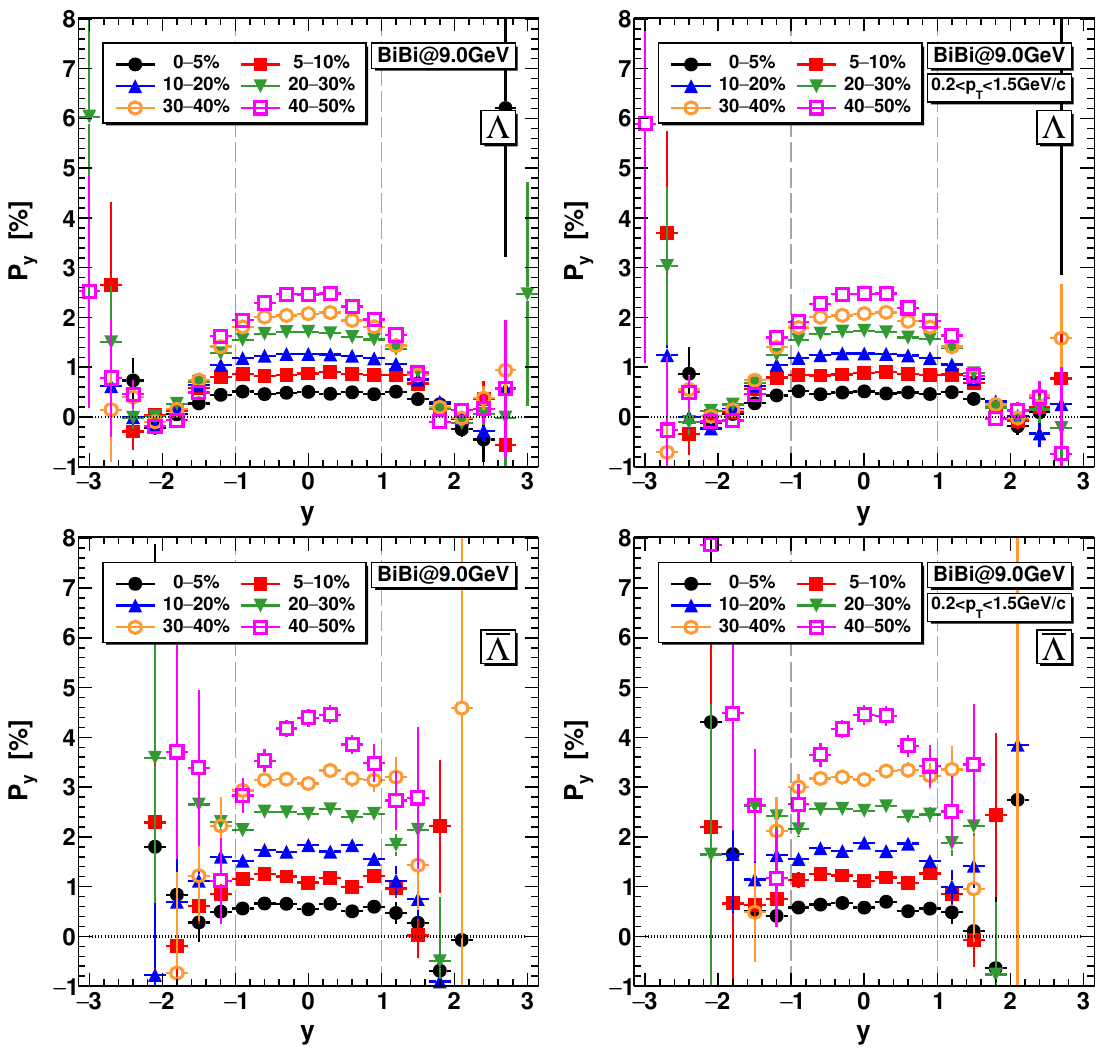}
	\caption{The spin polarizations of $\Lambda$ (upper row) and $\ALambda$ (lower row) hyperons as functions of the rapidity without momentum cuts (left panels) and with the transverse momentum cut $0.2<p_{T}<1.5\,$GeV$/c$ (right panels).}
\label{fig:pol-y}
\end{figure}

The polarization maps as shown in Fig.~\ref{fig:pol-map} can be prepared including only the events of a certain centrality class. Then, summing up vertically the $p_T$ bins we obtain the rapidity distributions of the polarization for different centrality classes that are shown in Fig.~\ref{fig:pol-y}. For centrality classes $<30\%$ the polarization weakly depends on rapidity if $|y|<1$ and increases if $|y|<0.5$ for classes 30--40\% and 40--50\%. This dependence is stronger for $\ALambda$. We observe also that the $p_T$ cut has a weak influence on the rapidity dependence of the spin polarization.

\begin{figure}
	\centering
\includegraphics[width=10cm]{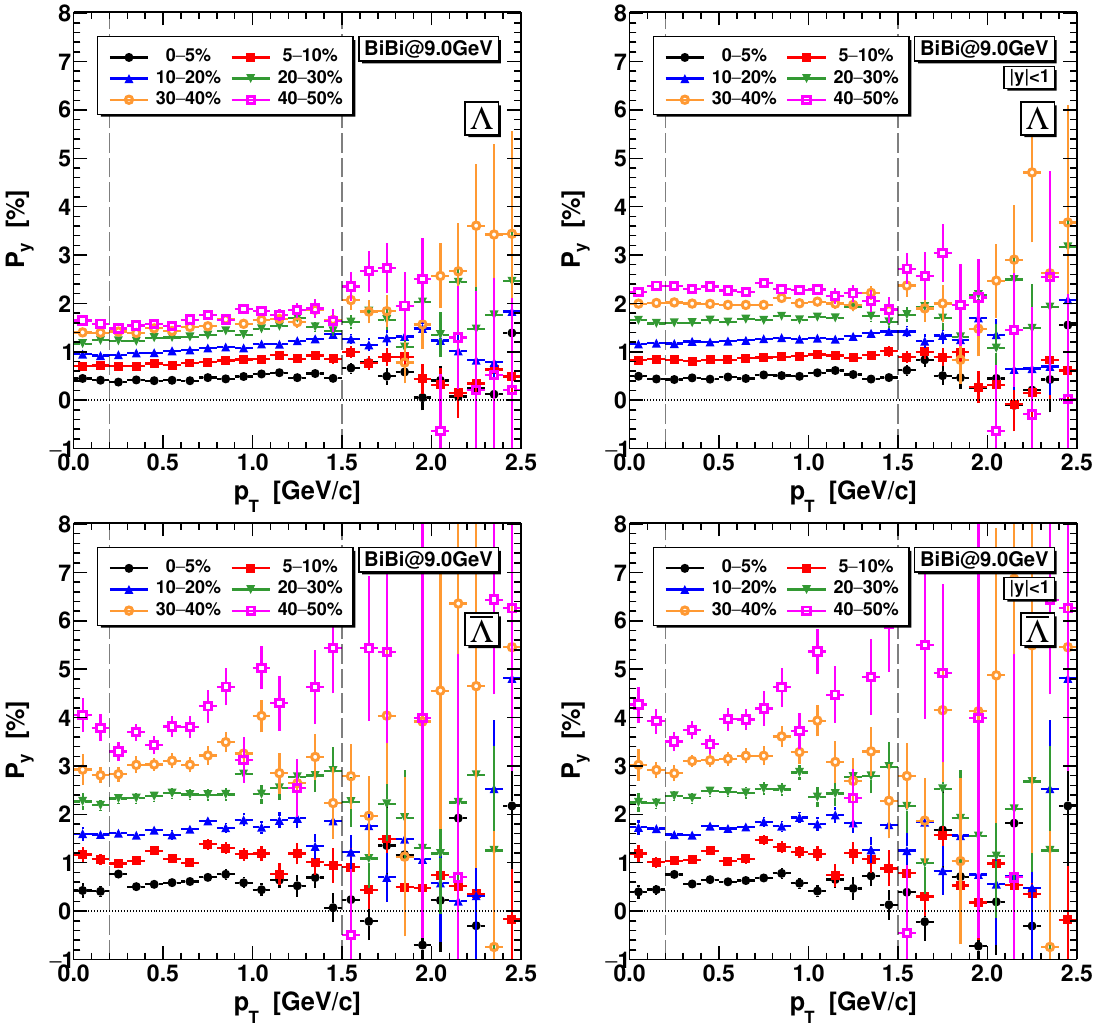}
\caption{The spin polarizations of $\Lambda$ (upper row) and $\ALambda$ (lower row) hyperons as functions of the transverse momentum without the rapidity cut (left panels) and with the rapidity cut $|y|<1$ (right panels).
}
\label{fig:pol-pt}
\end{figure}

In Fig.~\ref{fig:pol-pt} we show the results of the summation of the rapidity bins in Fig.~\ref{fig:pol-map} with the selection of different centrality classes.
We see that the spin polarization signal for $\Lambda$s almost does not change within the $0.2<p_T< 1.5$\,GeV window. For $p_T>1.5$\,GeV the polarization fluctuations increase significantly. For $\ALambda$ the polarization varies with $p_T$ stronger and for $p_T>1.5$\,GeV statistical errors become very large.
Comparing the left and right panels we conclude that the influence of rapidity cut is rather small.

Similar $y$ and $p_T$ dependence of the polarization is found for $\Sigma$, $\Xi$, and $\Omega$ hyperons and corresponding anti-particles.

\begin{figure}
	\centering
\includegraphics[width=10cm]{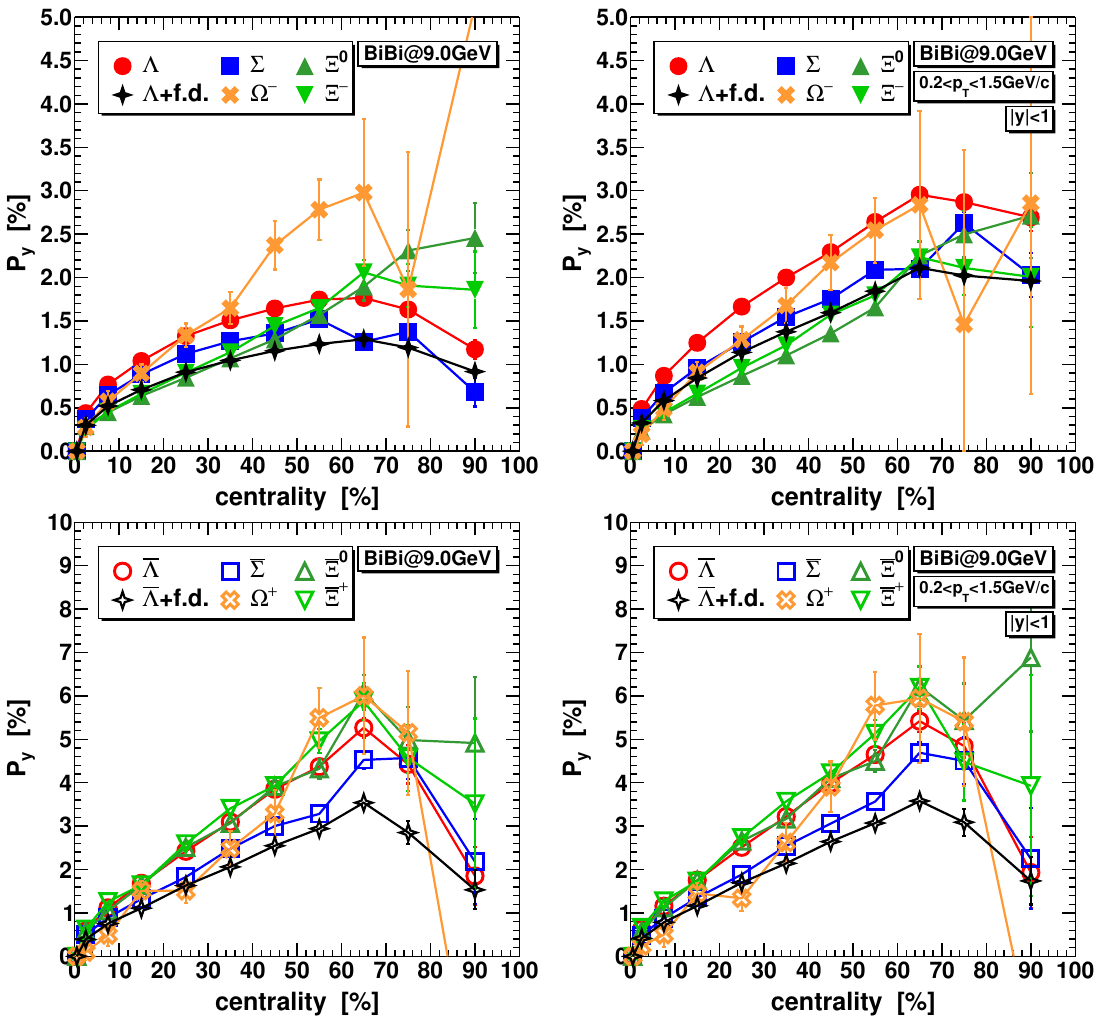}
	\caption{The dependence of the global spin polarization for various hyperons (upper row) and anti-hyperons (lower row) on the collision centrality for Bi+Bi collisions at $\sqrt{s_{NN}}=9.0\,$GeV without momentum cuts (left panels) and for the midrapidity region $|y|<1$ with the transverse momentum cut $0.2<p_{T}<1.5\,$GeV$/c$ (right panels).}
	\label{fig:pol-central}
\end{figure}

We turn now to the question of how the global spin polarization of various hyperon species changes with centrality.
It is shown in Fig.~\ref{fig:pol-central} without (left panel) and with acceptance constraints (right panel).
As seen, the polarization of all hyperons increases with a centrality class increase from the most central collision till the centrality class $\sim 60\mbox{--}70\%$. Then it starts decreasing. At the maximum, the $\Omega$ hyperons are most polarized in the case without rapidity and $p_T$ cuts. With the cuts, $\Lambda$s and $\Omega$s are similarly polarized. Polarization of all anti-hyperons are alike and maximal in the same centrality class as hyperons.
We see also that the polarization magnitude for anti-hyperons is approximately twice larger than that for hyperons. The acceptance cut almost does not affect the anti-hyperons, but significantly increases the resulting polarization for hyperons, especially for peripheral collisions.

As mentioned above, only strong decays of hyperon resonances are taken into account in the PHSD transport model. When all strong interactions stop, the family of $\Lambda$ hyperons is additionally populated due to weak and electromagnetic decays $\Xi\to\Lambda+\pi$ and $\Sigma^0\to \Lambda +\gamma$. The polarization transfer in these reactions is taken into account applying the relations from Refs.~\cite{Becattini-Karpenko-Lisa2017,KTV-PRC97}. In the same way, the feed-down corrections are applied for $\ALambda$s, see Refs.~\cite{Tsegelnik:2023isj,Voronyuk:2023vyu}. The $\Lambda$ and $\ALambda$ polarizations including the feed-down effects are shown in Fig.~\ref{fig:pol-central}. We observe a suppression of the $\Lambda$ and $\ALambda$ polarizations (up to $\approx36\%$), compare red and black symbols. The main role, thereby, is played by the decays of $\Sigma^0$($\overline{\Sigma}^0$) hyperons because of their large relative abundance.
Nevertheless, our calculations~\cite{Voronyuk:2023vyu} reproduce the polarization signals for both $\Lambda$ and $\ALambda$ at higher collision energy $\sqrt{s_{NN}}=11.5$\,GeV. The $\Lambda$ polarization for $\sqrt{s_{NN}}=7.7$\,GeV is also well described, however, the $\ALambda$ polarization is underestimated. The $\Xi (\AXi)$ and $\Omega(\AOmega)$ polarizations do not suffer from the feed-down effect.

\begin{table}
			\centering
\caption{The global spin polarizations and multiplicities of the $\Lambda$ and $\ALambda$ hyperons in Bi+Bi collisions at $\sqrt{s_{NN}}=9.0$\,GeV for the rapidity range $|y|<1$ and the transverse momentum range $0.2\,{\rm GeV}/c < p_T < 1.5\,{\rm GeV}/c$ depending on the different centrality classes. Numbers in brackets indicate statistical errors.
\label{tab:PolY-sNN}}
			\begin{tabularx}{0.96\linewidth}{ccccccccc}
                \toprule
& \multicolumn{2}{c}{$\Lambda$(prim.)} & \multicolumn{2}{c}{$\Lambda$} & \multicolumn{2}{c}{$\overline{\Lambda}$(prim.)} & \multicolumn{2}{c}{$\overline{\Lambda}$}\\
				\midrule
				centrality& $N$& $P_{y}, \%$ & $N$& $P_{y}, \%$ & $N$& $P_{y}, \%$  & $N$& $P_{y}, \%$\\
$ 10-50\%$ & 11.97(2) & 1.60(1) & 17.61(2) & 1.09(1) & 0.25(1) & 2.43(2)  & 0.48(1) & 1.62(1) \\
$ 10-60\%$ & 10.28(2) & 1.67(1) & 15.10(2) & 1.14(1) & 0.21(1) & 2.55(2)  & 0.40(1) & 1.70(1) \\
$ 10-70\%$ &  8.93(2) & 1.71(1) & 13.10(2) & 1.17(1) & 0.18(1) & 2.61(2)  & 0.34(1) & 1.74(1) \\
				\midrule
$ 20-50\%$ &  9.47(1) & 1.86(1) & 13.84(2) & 1.28(1) & 0.19(1) & 2.98(3)  & 0.35(1) & 1.99(2) \\
$ 20-60\%$ &  7.99(1) & 1.94(1) & 11.64(2) & 1.34(1) & 0.15(1) & 3.13(3)  & 0.29(1) & 2.08(2) \\
$ 20-70\%$ &  6.83(1) & 1.99(1) &  9.93(2) & 1.37(1) & 0.13(1) & 3.21(3)  & 0.24(1) & 2.14(2) \\
				\midrule
$ 30-50\%$ &  7.80(1) & 2.08(1) & 11.33(1) & 1.43(1) & 0.14(1) & 3.50(5)  & 0.27(1) & 2.32(3) \\
$ 30-60\%$ &  6.38(1) & 2.17(1) &  9.24(1) & 1.50(1) & 0.11(1) & 3.70(5)  & 0.21(1) & 2.44(3) \\
$ 30-70\%$ &  5.33(1) & 2.23(1) &  7.71(1) & 1.55(1) & 0.09(1) & 3.81(5)  & 0.17(1) & 2.52(3) \\
				\midrule
$ 40-50\%$ &  6.27(1) & 2.25(2) &  9.06(1) & 1.56(1) & 0.11(1) & 3.99(9)  & 0.20(1) & 2.63(6) \\
$ 40-60\%$ &  4.90(1) & 2.37(2) &  7.06(1) & 1.65(1) & 0.08(1) & 4.22(8)  & 0.15(1) & 2.78(5) \\
$ 40-70\%$ &  4.00(1) & 2.45(2) &  5.74(1) & 1.71(1) & 0.06(1) & 4.35(8)  & 0.12(1) & 2.87(5) \\
				\midrule
$ 50-70\%$ &  2.86(1) & 2.66(3) &  4.08(1) & 1.87(2) & 0.04(1) & 4.80(13) & 0.08(1) & 3.17(8) \\
$50-100\%$ &  1.54(1) & 2.67(3) &  2.18(1) & 1.89(2) & 0.02(1) & 4.60(13) & 0.04(1) & 3.06(8) \\
				\midrule
$60-100\%$ &  1.04(1) & 2.79(5) &  1.46(1) & 2.00(3) & 0.01(1) & 4.57(24) & 0.02(1) & 3.09(15)\\
$70-100\%$ &  0.66(1) & 2.71(8) &  0.91(1) & 1.94(6) & 0.01(1) & 3.72(44) & 0.01(1) & 2.55(27)\\
				\bottomrule
			\end{tabularx}
		\end{table}

The experimental feasibility of polarization signal registration depends on the particle abundance~\cite{Nazarova:2024jic}. The magnitude of the polarization signal is compared with the number of produced hyperons for different centrality classes in Table~\ref{tab:PolY-sNN}.
The largest number of particles can be registered for the broadest centrality classes 10--50\% and 10--60\%. Narrowing the centrality window towards the less-central collision gives a larger global spin polarization, e.g., 2.37\% for primary $\Lambda$s and 4.22\% for primary $\ALambda$s, but the number of produced hyperons decreases by a factor of $\sim 3$.
The account for the hyperon feed-down increases the number of $\Lambda$s by a factor of $1.5$ and the $\ALambda$ number by a factor of $\sim 2$. At the same time, the global spin polarization falls down equally for $\Lambda$s and $\ALambda$s by a factor of $\sim 1.5$.

At hand of Table~\ref{tab:PolY-sNN} we may conclude that is more feasible to measure the polarization signal in semi-central collisions or in broad centrality bins. For example, considering the $50-70$\% centrality bin instead of the $10-50\%$ bin we may increase polarization signals for $\Lambda$s by 60\% and for $\ALambda$ by a factor of 2, but the number of particles drops by a factor of $\sim 4$ for $\Lambda$s and by a factor of $\sim 6$ for $\ALambda$s.

\section{Angular dependence of the spin polarization}\label{sec:correlat}

\begin{figure}
	\centering
\includegraphics[width=11cm]{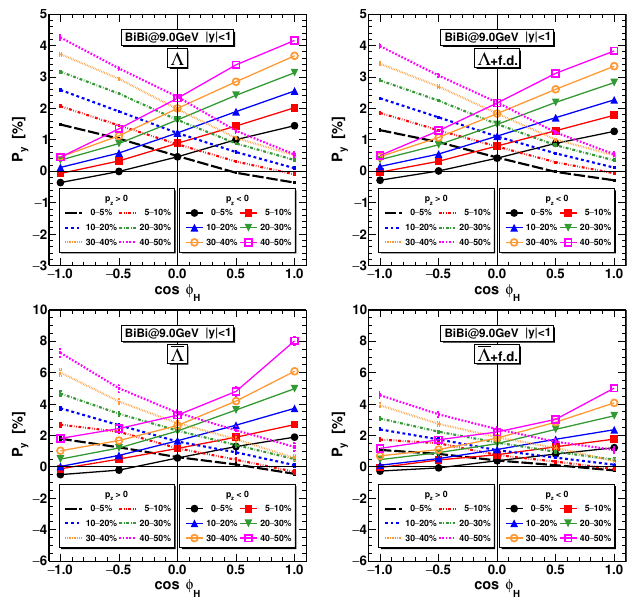}
	\caption{The spin polarizations of $\Lambda$ (upper row) and $\ALambda$ (lower row) hyperons as functions of the azimuthal angle in the transverse momentum plane for various centrality classes without feed-down due to electromagnetic and weak decays (left panels) and with it (right panels).
}
\label{fig:pol-V1-y}
\end{figure}

Several sources of the spin polarization in heavy-ion collisions are considered in the literature. Some of them are related to the vorticity of the medium~ \cite{Becattini-Chandra2013}, others to the formation of helicity field, $h=(\vec{\omega} \vec{v})$,~\cite{Rogachevsky-ST-2010,BGST-Hseparation}.
It is interesting whether experimental data could reveal information about an operative polarizing mechanism.
In Ref.~\cite{helicity:2022} we argue that the helicity field created in HICs has a prominent plane symmetry: on one side of the reaction plane, the helicity is positive and is negative on the other side. Also, we demonstrated that, by selecting hyperons with a positive or negative projection of the hyperon momentum on the axis perpendicular to the reaction plane, one can probe regions with different helicities. The mechanism of the spin polarization, which we consider here, is due to the vorticity field. Our calculations show that the latter has a peculiar structure of two asymmetrical rings (bubliks) moving in opposite directions, see Fig.~\ref{fig:V-W-fields}. One can try to separate the hyperons stemming from the most vortical parts of the bubliks. They would have opposite polarization directions, $P_y$. For this, we select positive or negative projections of the momenta along the collision axis, $p_z\gtrless 0$, and look at the dependence of the hyperon polarization on the azimuthal angle, $\cos\phi_H=p_x/p_T$. The spin polarization as a function of $\cos\phi_H$ is shown in Fig.~\ref{fig:pol-V1-y} for $p_z>0$ and $p_z<0$. We observe a drastically different behavior of the $P_y(\cos\phi_H)$ function: it is rising for $p_z>0$ and decreasing for $p_z<0$.
It can be parameterized as
\begin{align}
\langle P_{y,H}\rangle\propto \big(1+ \beta_{H,1}(p_z)\cos\phi_{H} + \beta_{H,2}(p_z)\cos^2\phi_H\big), \quad H=\Lambda\,,\,\, \Xi\,,\,\,\ALambda\,,\,\,\AXi\,,
\end{align}
where the coefficient $\beta_{H,1}(p_z)$ is approximately anti-symmetrical function of $p_z$, and $\beta_{H,2}(p_z)$ is approximately symmetrical one. For the given statistics, $|\beta_{2,H}|$ is much smaller than $|\beta_{1,H}|$ for hyperons $\Lambda$ and $\Xi$ but for anti-hyperons $\ALambda$ and $\AXi$ its contribution is visible, see plots in lower panels of Fig.~\ref{fig:pol-V1-y} and in the right panel of Fig.~\ref{fig:pol-V1-y-xi}.
For primary $\Lambda$s and $\ALambda$s without the electromagnetic and weak feed-down the angular dependence of the polarization signal is more pronounced as seen in the left panels of Fig.~\ref{fig:pol-V1-y}.

The experimental observation of such patterns would speak for the vortical mechanism of polarization and the particular structure of the vorticity field.

\section{Conclusions}\label{sec:concl}

The hyperon production in collisions of two bismuth nuclei at energy $\sqrt{s_{NN}}=9.0$\,GeV is studied within the PHSD transport model. Yields of strange particles, $\Lambda$, $\Sigma$, $\Xi$, and $\Omega$ and corresponding anti-particles are calculated for various centrality classes. The hierarchy of $\Lambda$, $\Xi$, and $\Omega$ yields for the most central collisions is found to be similar to the hierarchy observed in central Pb+Pb collisions at $\sqrt{s_{NN}}=8.77$\,GeV by the NA49 collaboration. Also, the hierarchy of measured anti-particles $\ALambda$ and $\AXi$ is alike the experimental one. The influence of the transverse momentum, $p_T$, and rapidity, $y$, cuts is investigated. The hyperon yields demonstrate the stronger dependence on the rapidity cut than on the $p_T$ cuts. The rapidity spectra of $\Lambda$ and $\Xi$ hyperons and corresponding anti-particles are calculated and analyzed. The $y$ distributions for anti-hyperons are found to be more compact than those of hyperons. The slope parameters of the $p_T$ spectrum for $\Lambda$ and $\Xi$ show a weak variation with the collision centrality, which, however, is significantly stronger for $\ALambda$ and $\AXi$. Restricting the rapidity range leads to a significant increase of the slope parameter: by $\sim 15$\,MeV for $\Lambda$s and by $\sim  30$\,MeV for $\Xi$s. Changes in the slopes for anti-hyperons are less significant.

\begin{figure}
	\centering
	\includegraphics[width=11cm]{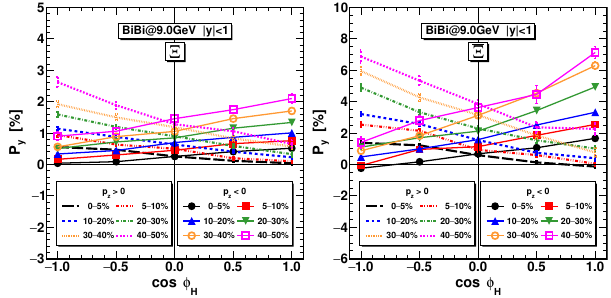}
	\caption{The spin polarizations of $\Xi$ (left panel) and $\AXi$ (right panel) hyperons as functions of the azimuthal angle in the transverse momentum plane for various centrality classes.
	}
	\label{fig:pol-V1-y-xi}
\end{figure}

We performed fluidization of the particle distributions generated by the PHSD transport code and determined evolutions of the velocity and vorticity fields assuming that the velocity is defined within the Landau frame. Vorticity is shown to be concentrated in the form of two asymmetrical vortex rings moving in opposite directions. Averaged spin polarizations of hyperons and anti-hyperons induced by the local thermal vorticity are evaluated. We analyzed the dependence of the polarization signal on the collision centrality and momentum cuts. For anti-hyperons, the $p_T$ and rapidity cuts do not change the polarization signal for centralities $<40\%$. However, the cuts affect the polarizations of hyperons. The polarization signal is almost linearly rising with the centrality (from central to peripheral collisions) exhibiting a maximum for the 60--$70\%$ centrality class. The final polarization of $\Lambda$ and $\overline{\Lambda}$ particles are calculated with the account for feed-down effects due to the weak, $\Xi\to \Lambda +\pi$, and electromagnetic, $\Sigma^0\to \Lambda + \pi^0$, decays. The feed-down increases the number of final $\Lambda$s by a factor of 1.5 and $\ALambda$ by a factor of 2. The polarization signal decreases equally for $\Lambda$s and $\ALambda$s by a factor of $\sim 1.5$. We showed that the vortical mechanism of the spin polarization and the particular structure of the vorticity field manifest themselves through the dependence of the polarization signal on the azimuthal angle of the outgoing hyperon. Thereby, the polarization increases for particles with $p_z>0$ and decreases for $p_z<0$, see Fig.~\ref{fig:pol-V1-y}. Also, for $\Lambda$ hyperons, the dependence is almost linear, whereas for $\ALambda$ there is a small admixture of a quadratic component. For primary $\Lambda$ and $\ALambda$ hyperons the angular dependence is more pronounced. A similar pattern is found also for the $\Xi$ and $\AXi$ hyperons as shown in Fig.~\ref{fig:pol-V1-y-xi}.
Thus, we conclude that the experimental study of the angular dependence of the polarization signal in the transverse plane could reveal information about possible polarization mechanisms.

\vspace{6pt}



\authorcontributions{All authors have equally contributed to the research and preparation of the publication.}

\funding{This research was partially supported by the Slovak grant VEGA~1/0521/22.
The calculations were performed on the ``Govorun'' computational cluster provided by the Laboratory of Information Technologies of JINR, Dubna.
}

\institutionalreview{Not applicable}

\informedconsent{Not applicable}

\dataavailability{No explicit data set are prepared.}

\acknowledgments{We thank D.N.~Voskresensky, Yu.B.~Ivanov and A.V. Taranenko for discussions.}

\conflictsofinterest{The authors declare no conflict of interest.}



\abbreviations{Abbreviations}{
The following abbreviations are used in this manuscript:\\

\noindent
\begin{tabular}{@{}ll}
AGS &  Alternating Gradient Synchrotron\\
BM$@$N &  Baryonic Matter at Nuclotron\\
EoS & Equation of State\\
HADES &  High Acceptance Di-Electron Spectrometer  \\
HIC & Heavy-Ion Collision\\
LHC & Large Hadron Collider\\
MPD & Multi-Purpose Detector\\
NICA & Nuclotron-based Ion Collider fAcility\\
PHSD & Parton Hadron String Dynamics\\
QGP & Quark-Gluon Plasma \\
QGSM & Quark-Gluon String Model\\
RHIC & Relativistic Heavy Ion Collider  \\
SIS & Schwer-Ionen-Synchrotron (heavy-ion synchrotron) \\
SPS & Super Proton Synchrotron\\
STAR & Solenoidal Tracker at RHIC \\
UrQMD & Ultra-relativistic Quantum Molecular Dynamics
\end{tabular}
}

\begin{adjustwidth}{-\extralength}{0cm}

\reftitle{References}



\bibliography{MPD-flow-polar.bib}

%


\PublishersNote{}
\end{adjustwidth}
\end{document}